# Experimental Validation of Coherent Joint Transmission in a Distributed-MIMO System with Analog Fronthaul for 6G

Rafael Puerta[1,2], Mahdieh Joharifar[2], Mengyao Han[2], Anders Djupsjöbacka[3], Vjaceslavs Bobrovs[4], Sergei Popov[2], Oskars Ozolins[2,3,4], and Xiaodan Pang[2,3,4]

[1]Ericsson Research, Ericsson AB, Stockholm, Sweden
[2]Department of Applied Physics, KTH Royal Institute of Technology, Stockholm, Sweden
[3]Network Unit, RISE Research Institutes of Sweden, Stockholm, Sweden
[4]Institute of Telecommunications, Riga Technical University, Riga, Latvia
rafael.puerta@ericsson.com, {rafaelpr, mahdieh, mengyaoh, ozolins, sergeip, xiaodan}@kth.se

*Abstract*—The sixth-generation (6G) mobile networks must increase coverage and improve spectral efficiency, especially for cell-edge users. Distributed multiple-input multiple-output (D-MIMO) networks can fulfill these requirements provided that transmission/reception points (TRxPs) of the network can be synchronized with sub-nanosecond precision, however, synchronization with current backhaul and fronthaul digital interfaces is challenging. For 6G new services and scenarios, analog radio-over-fiber (ARoF) is a prospective alternative for future mobile fronthaul where current solutions fall short to fulfill future demands on bandwidth, synchronization, and/or power consumption. This paper presents an experimental validation of coherent joint transmissions (CJTs) in a two TRxPs D-MIMO network where ARoF fronthaul links allow to meet the required level of synchronization. Results show that by means of CJT a combined diversity and power gain of +5 dB is realized in comparison with a single TRxP transmission.

## I. Introduction

International Mobile Telecommunications (IMT) towards 2030, i.e., future sixth-generation (6G) mobile networks, are expected to bring several improvements over previous generations. 6G promises to offer much higher speeds and capacity compared to fifth-generation (5G) mobile networks, allowing for faster data transmission and improved connectivity. This means that users will be able to enjoy a more seamless experience when connected to 6G mobile networks, with reduced latency and buffering times [1]. Another key advantage of 6G technology is that it will support a larger number of devices and connections. This is important for industries such as the future industrial internet of things (IIoT), where there is a growing need for a vast number of devices to be connected to the internet. 6G will be able to accommodate these devices and allow for their seamless integration into the connected world [2]. While 6G technology holds great promise, there are still many challenges that need to be overcome in order for it to become a reality [3]. Among the many enablers of 6G mobile networks, multiple-input multiple-output (MIMO) technology continues to play a major role. Today, MIMO systems still underperform in comparison with the theoretical channel capacity bounds, thus, new approaches are critical to bridge this gap. Solutions to improve current MIMO technology solutions include implementing much larger antenna arrays, enhanced hybrid beamforming, distributed deployments, artificial intelligence (AI), among others [4].

Distributed MIMO (D-MIMO) is a wireless technology that has gained significant attention in recent years [5], [6] as a highly effective solution for increasing the capacity and coverage of wireless networks that can fulfill 6G new services and scenarios foreseen requirements, e.g., extreme high data rate and improved coverage [1]. In D-MIMO, multiple antennas are deployed at different positions within a network, such as at different base stations or access points, where the antennas can work jointly, synchronously, and coherently to improve performance and serve multiple users simultaneously [6]. This approach is different from traditional MIMO systems, where all antennas are co-located at a single location of a cell. By distributing the antennas, D-MIMO can improve the spatial multiplexing gain, reduce interference, and increase the spectral efficiency of the network. This means that D-MIMO can support more users, transmit data faster and more reliably, and provide better coverage in challenging environments improving the quality of service for users located at the cell-edge. Thus, D-MIMO can reduce power consumption by moving the network closer to the users and can increase the data rates and coverage since is highly likely that at least one antenna is in the proximity of each user with an unobstructed view reducing transmission losses. Additionally, besides conventional wired links, D-MIMO transmission/reception points (TRxPs) can be connected via wireless or free space optics (FSO) backhaul and/or fronthaul links as in a relay network [7]–[9] providing flexibility and scalability.

In general, a D-MIMO network performs better than a conventional co-located system, however new challenges arise due to additional requirements on bandwidth, synchronization, and/or power consumption [3], [5], thus novel solutions are needed to realize the potential of D-MIMO in a cost-effective manner. To take full advantage of the diversity and power gains of D-MIMO networks, i.e., coherent joint transmissions (CJTs) from two or more TRxPs, stringent synchronization requirements with sub-nanosecond precision must be met. An appealing solution to realize CJTs is to use analog fronthaul links in combination with centralized processing. Analog radio-over-fiber (ARoF) links are an alternative that can fulfil the preceding requirements facilitating the deployment of D-MIMO networks. ARoF links have been proven to fulfill the 3rd Generation Partnership Project (3GPP) [10] radio frequency (RF) mandatory error vector magnitude (EVM) and adjacent channel leakage power ratio (ACLR) transmitter requirements for the frequency range 1 (FR1) [11], validating ARoF links as a feasible solution for future deployments such as D-MIMO.

In this paper, we experimentally validate CJTs in a two TRxPs D-MIMO network. Centralized signal processing and ARoF fronthaul links allow to synchronize both TRxPs. Synchronization is possible since channel estimation includes the phase and amplitude change due to both the wired path (i.e., fronthaul link) and wireless path of each TRxP providing all the necessary information for CJT precoding. To the best of our knowledge, under our setup test conditions, this is the first time that CJT is realized where ARoF fronthaul technology is used. The remainder of this paper is organized as follows: Section II provides CJT and the system model background information. Section III and Section IV describe the experimental setup and experimental results, respectively. Finally, conclusions are provided in Section V.

## II. COHERENT JOINT TRANSMISSION

Joint transmission (JT) techniques are used for the simultaneous transmission from multiple TRxPs to the same user. A scheme used in coordinated multi-point (CoMP) technology is joint processing including JT and joint reception. JT includes two methods: non-coherent joint transmission (NCJT) and CJT. With the former, the network does not rely on explicit channel information and no RF phase coherency can be realized. Thus, the gain that can eventually be achieved is only power gain, i.e., that the power of several TRxPs is used to serve the same user. On the other hand, with CJT, the explicit channel information between the user and two or more TRxPs is used to compute the transmission precoder of the TRxPs involved in the JT. In practice, with CJT the largest MIMO gains can be achieved, namely, diversity and power gains. However, CJT has challenging requirements on the relative phase coherency and time synchronization of the cooperating TRxPs increasing the complexity of its implementation.

Assuming time division duplex (TDD) operation, where channel reciprocity applies, an alternative to achieve CJT in a D-MIMO network is to perform channel estimation and precoding in each TRxP and to exchange this information between the cooperating TRxPs for synchronization [3], [10]. Some issues with this approach are the additional overhead due to this information exchange and that each TRxP must perform significant digital signal processing (DSP) increasing its power consumption and complexity. Conversely, if the DSP is centralized, the complexity problem can be alleviated, however, synchronization is still challenging.

In a D-MIMO network with centralized processing, the TRxPs can be connected via wired or wireless fronthaul links [7]–[9]. Depending on the deployment characteristics and needs, the length of these links is different, and this difference may span from few to hundreds of meters adding delays in the transmission from each TRxPs. To fulfill CJT requirements, fronthaul links delays must be considered for synchronization with sub-nanosecond precision, however, with conventional digital fronthaul interfaces, this is challenging to achieve, and usually additional timing references or protocols may be required, e.g., global navigation satellite system (GNSS) technology, precision time protocol (PTP) [13]. Thus, if the different TRxPs are not fully synchronized and their transmissions are not phase aligned, it is often the case that only NCJTs is feasible. Analog fronthaul is an appealing solution for D-MIMO fronthaul [10], [14]. ARoF fronthaul links technology does not use bits to transmit information since the optical carriers that propagate through the fiber are modulated with RF signals, i.e., ARoF signals, allowing to estimate the amplitude and phase changes of the wireless channel and fronthaul links.

### A. System Model

A simple multiple-input single-output (MISO) system is considered where the D-MIMO network comprises N TRxPs and the receiver is a single mobile station (MS). All TRxPs and the MS are equipped with a single antenna. In the downlink direction, the baseband received signal at the MS can be mathematically expressed as:

$$y = \mathbf{h}\mathbf{w}s + n \quad (1)$$

where $\mathbf{h} = [h_1, h_2, \ldots, h_N]$ is the 1×N channel vector assumed to be flat-fading, s is the transmitted data symbol, $\mathbf{w}$ is the N×1 precoding weights vector, and n is the additive white Gaussian noise (AWGN) scalar distributed as $\mathcal{CN}(0, N_0)$ where $N_0$ is the noise power spectral density. The largest signal-to-noise ratio (SNR) gain can be achieved by computing the precoder $\mathbf{w}$ in such a way that the signals of the various TRxPs add up in-phase (coherently), e.g., maximum ratio transmission (MRT) precoding. Thus, the AWGN channel received SNR can be expressed as [15]:

$$\text{SNR} = \frac{P\|\mathbf{h}\|^2}{N_0} = \sum_{i=1}^{N} \frac{p_i |h_i|^2}{N_0} \quad (2)$$

where $p_i$ is the average transmit power per TRxP. For example, for the particular case of 2 TRxPs with equal transmit power and channels magnitude ($|h_1| = |h_2|$), in comparison with a single TRxP transmission, the SNR theoretical gain is:

$$G = 10 \log_{10}\left(\frac{2p \cdot 2|h_1|^2}{p|h_1|^2}\right) = 6.02 \text{ dB} \quad (3)$$

Note that, since the TRxPs are not co-located as in conventional MIMO systems, a power constrain is not applied and therefore the transmit power is doubled when 2 TRxPs are assumed.

## III. EXPERIMENTAL SETUP

### A. Hardware

The experimental setup block diagram used in our tests is depicted in Fig. 1. In our setup, TRxP 1 is connected via an ARoF fronthaul link while TRxP 2 fronthaul consists only of a 1.2 m coxial cable. A 10 MHz orthogonal frequency-division multiplexing (OFDM) signal with a subcarrier spacing (SCS) of 30 kHz, 24 resource blocks (RBs), and a carrier frequency of 622 MHz is used for all tests. Although not a TDD band, this New Radio (NR) band (n71) was chosen provided that an anechoic chamber was not available for the tests and instead a narrow band filter at this frequency was available.

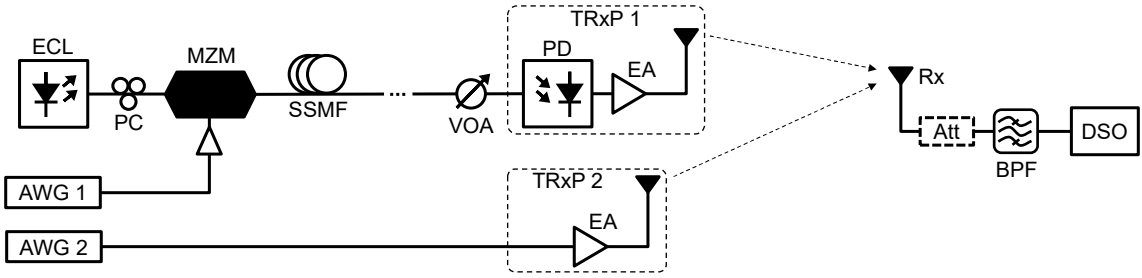

Fig. 1. Experimental setup block diagram.

To transmit the ARoF signal to TRxP 1, an external cavity laser (ECL) with a wavelength of 1550.9 nm, a linewidth of less than 100 kHz, and an output power of 12 dBm, serves as the light source. An arbitrary waveform generator (AWG) with 10-bit vertical resolution and a sampling frequency of 50 GSa/s generates the OFDM analog signals. These analog signals modulate the ECL output via a single-drive Mach-Zehnder modulator (MZM) with a Vπ equal to 4.2 V, generating ARoF signals by biasing the MZM close to the quadrature point. The light polarization at the MZM input is adjusted by a polarization controller (PC). The MZM output is transmitted through standard single-mode fiber (SSMF), and the received optical power is set by a variable optical attenuator (VOA) before optical heterodyne detection in a photodetector (PD) with an analog bandwidth of 9 GHz and a responsivity of 0.8 A/W. The signal is then amplified by an electrical amplifier (EA) with a gain of 22 dB and a noise figure (NF) of 5.5 dB. Similarly, the electrical signal transmitted to TRxP 2 is amplified by an EA with a gain of 30 dB and a NF of 6 dB. For wireless transmission, we used multiband antennas with a peak gain of approximately 3 dBi. The selected EAs provide a similar performance on both TRxPs provided their different fronthaul implementations.

After wireless transmission, the received signal is first filtered with a narrow bandpass filter (BPF) with a central frequency of 622 MHz. For CJT tests, we optionally use an electrical attenuator before the BPF to adjust the received signal power in order to validate the CJT power and diversity gains. In addition, note that we adjusted the geometry of our setup to achieve nearly flat fading conditions on the received signals from both TRxPs to realize additive white Gaussian noise (AWGN) channels, however, it was difficult to retain the same distance between the antennas, thus, the distances between the receiver antenna and the TRxP 1 and TRxP 2 antennas are 1 m and 1.3 m respectively. Furthermore, the distance between the TRxPs antennas is around 3 lambdas, namely, ~1.5 m with a carrier frequency of 622 MHz. After bandpass filtering, the received signals are stored in a digital storage oscilloscope (DSO) for further offline signal processing. Fig. 2 displays a photograph of the experimental setup.

### B. Receiver Signal Processing and Channel Estimation

After analog to digital conversion, first a digital Costas loop is used for carrier recovery and subsequent digital down conversion (DDC) to baseband [16], [17]. Then, after cyclic prefix (CP) removal and time domain to frequency domain conversion by the fast Fourier transform (FFT), a zero-forcing (ZF) equalizer is computed where the equalizer complex-valued coefficients at each reference symbol subcarrier are averaged using a moving window with a maximum length of 19. The ZF equalizer frequency domain coefficients before averaging are given by:

$$\alpha(t,f)e^{i\varphi(t,f)} = \frac{Z(t,f)}{I(t,f)} \quad (4)$$

where $I(t,f)$ are the post-FFT ideal reference symbols and $Z(t,f)$ are the post-FFT received symbols. The complete equalizer coefficients are calculated, averaged, and interpolated as indicated in [11] to follow 3GPP conformance test methods. Then, the equalized symbols are compared with the ideal reference symbols to determine the received reference signal EVM.

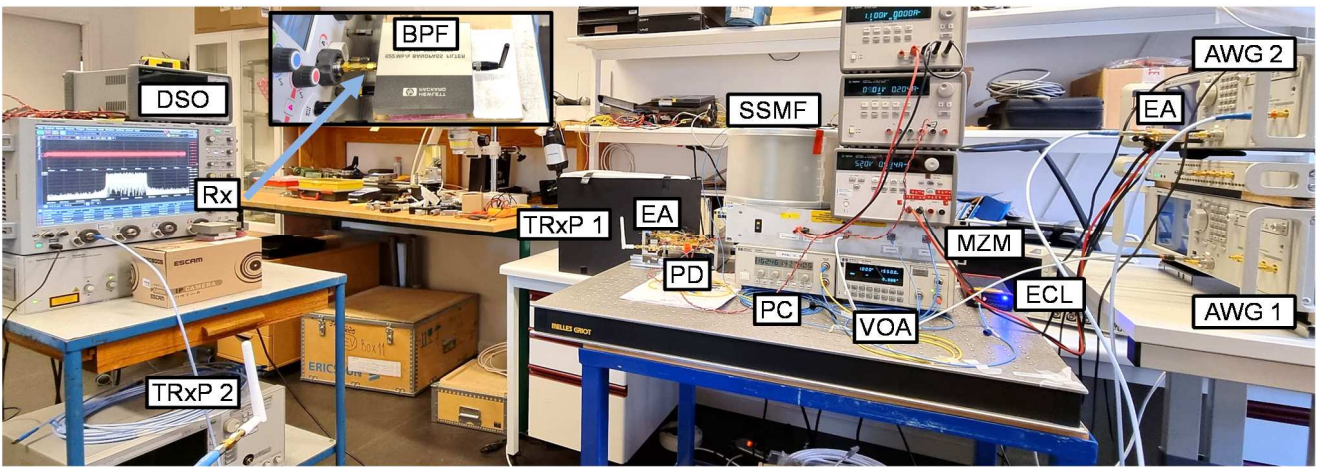

Fig. 2. Photo of experimental setup.

For channel estimation, although in practice reciprocity-based schemes are preferred, due to limitations in hardware, we estimated the channel at the receiver and used the explicit channel estimation information at the transmitter. Therefore, the results in this work can be understood as nearly ideal where perfect calibration between the uplink and downlink front-ends chains is assumed. The least squares estimator (LSE) is used to calculate the channel estimate. To determine the estimates of both TRxPs, transmission of the same signal with each TRxP is done separately. The instantaneous channel estimate in the frequency domain per subcarrier is computed as follows:

$$\hat{\mathbf{H}} = \mathbf{Y}\mathbf{X}^H(\mathbf{X}\mathbf{X}^H)^{-1} \quad (5)$$

where $\mathbf{X}$ is a diagonal matrix with known symbol sequences, $\mathbf{Y}$ is the time aligned sequences after wireless transmission, and $\hat{\mathbf{H}}$ is the channel estimate. It is noted that $\hat{\mathbf{H}}$ includes the amplitude and phase changes due to the wireless channel, analog front-ends, and fronthaul links. Subsequently, a digital lowpass band filter is used to clean the channel estimates as shown in Fig. 3. Finally, to realize CJT, only TRxP 2 signal is time aligned and precoded. The precoder is synthesized using the complex ratio $\hat{\mathbf{h}}_1/\hat{\mathbf{h}}_2$ where $\hat{\mathbf{h}}_1$ and $\hat{\mathbf{h}}_2$ corresponds to the subcarriers vector channel estimates of TRxP 1 and TRxP 2 respectively. Thus, the OFDM signal transmitted by TRxP 2 is time and phase aligned to the signal transmitted by TRxP 1.

## IV. EXPERIMENTAL RESULTS

First, we assessed the end-to-end EVM performance of TRxP 1 including the ARoF fronthaul link. By optimizing the operating points of the different optical and electrical devices, an EVM root-mean square (RMS) value of 1.91% is achieved after ZF equalization when all subcarriers carry high-order 256-QAM symbols as shown in Fig. 4. Under these conditions, the PD received optical power is -11 dBm.

To test D-MIMO CJT, an electrical attenuator is added before the receiver's BPF to worsen the TRxPs transmissions performance. Then, for a fair assessment of CJT, the EVM performance of both TRxPs is adjusted to be nearly the same when transmitting separately. Additionally, to test the flexibility of D-MIMO with analog fronthaul links, TRxP 1 fronthaul link fiber length is set to 800 m, and since TRxP 2 fronthaul is 1.2 m of coaxial cable, the delay difference between fronthaul links is ~3.8055 μs (with the approximation that light propagates in a fiber at 70 percent the speed of light in vacuum) which is larger than the OFDM signals CP duration with 30 KHz SCS, i.e., 2.34 μs [11]. Consequently, signal precoding alone is not enough to realize CJT, and additional delay difference measurement and time alignment are necessary. Following the preceding conditions, Fig. 5 and Fig. 6 show the received symbols constellations after separate wireless transmissions of TRxP 1 and TRxP 2 respectively. The EVM RMS values are 9.14% and 9.27% which are nearly the same. Furthermore, the measured delay difference between fronthaul links is 3.9062 μs which is very close to theoretical value of 3.8055 μs, where the difference is due to the additional delay of the various optical and electrical devices. Subsequently, as indicated in Section III.B, channel estimation is done separately for TRxP 1 and TRxP 2 and these channel estimates are used to calculate the precoder for TRxP 2 signal. Lastly, the precoded signal is time aligned in accordance with the measured delay difference between fronthaul links. Fig. 7 and Fig. 8 show the received symbols constellations with and without precoding and time alignment when both TRxPs are transmitting simultaneously. As expected, if no CJT processing is done, an EVM of 99.74% is measured since no coherency and inter-symbol interference (ISI) occurs even with OFDM CP insertion. Conversely, if precoding and time alignment is applied, an EVM of 5.01% is achieved since the transmitted signals are time and phase aligned maximizing the network diversity and power gains.

Provided that the TRxPs transmission channel conditions are close to flat fading, to determine the D-MIMO combined diversity and power gains, we use the following relationship between the EVM and the SNR [18]:

$$\text{EVM}^2_{\text{RMS}} \approx \frac{1}{\text{SNR}} \quad (6)$$

Thus, we calculate that the CJT gain is ~5.35 dB in comparison with TRxP 2 transmissions alone, and ~5.22 dB in comparison with just TRxP 1 transmissions alone. Note that for a 2 TRxPs D-MIMO network, the theoretical CJT gain is 6 dB. One reason why the experimental CJT gain is lower, is due to the fact that the EVM performance of both TRxPs was set equal even though the wireless distances between antennas is not the same, therefore the transmit power of TRxPs is not the same. In addition, the antennas of both TRxPs are not identical and our wireless transmissions do not experience a perfect flat-fading channel.

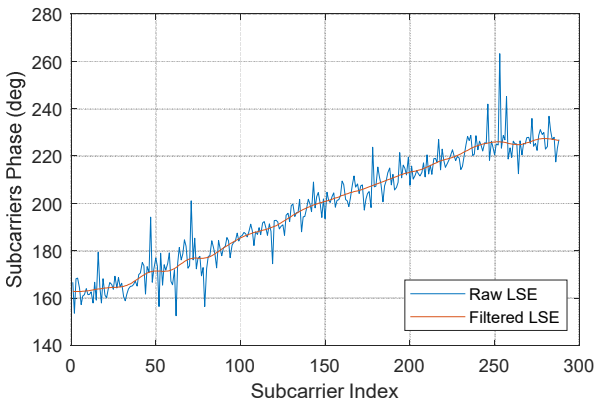

Fig. 3. TRxP 1 example of an OFDM symbol channel estimate with 24 RBs (288 subcarriers) before and after filtering.

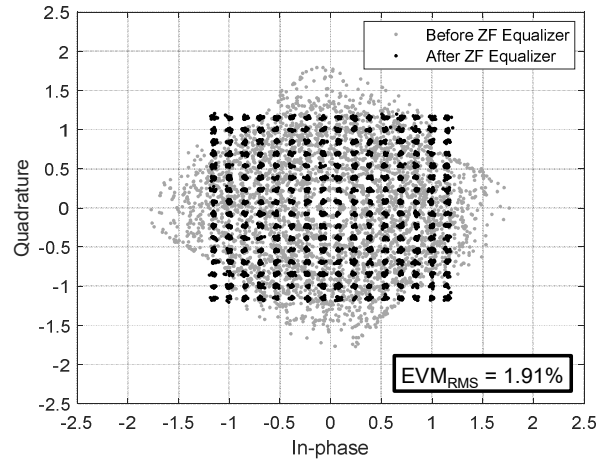

Fig. 4. TRxP 1 optimal received symbols constellation and EVM measurement before and after ZF equalization.

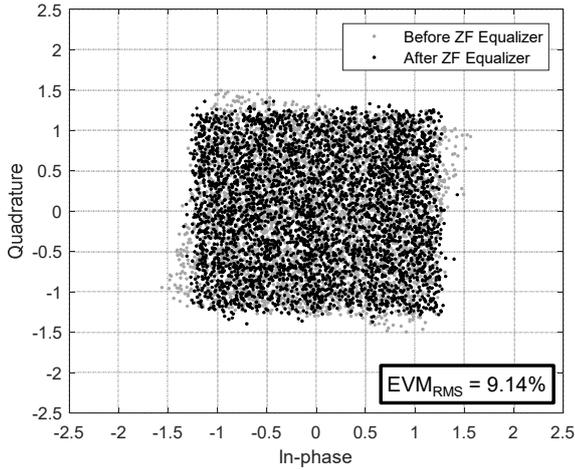

Fig. 5. TRxP 1 received symbols constellation and EVM measurement before and after ZF equalization.

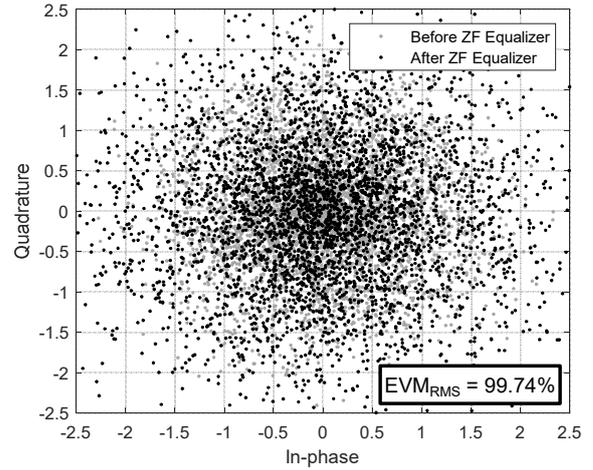

Fig. 7. TRxPs transmitting simultaneously. NCJT received symbols constellation and EVM measurement before and after ZF equalization.

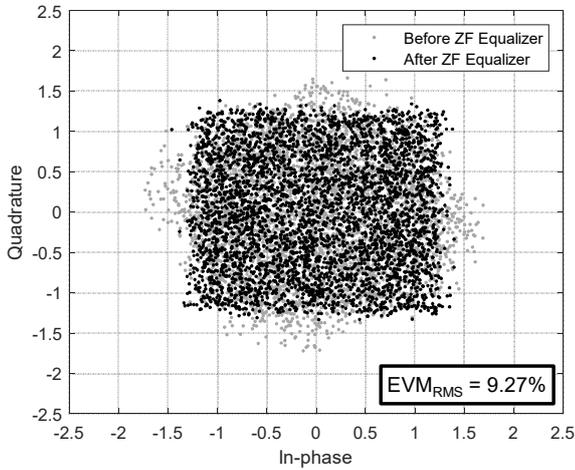

Fig. 6. TRxP 2 received symbols constellation and EVM measurement before and after ZF equalization.

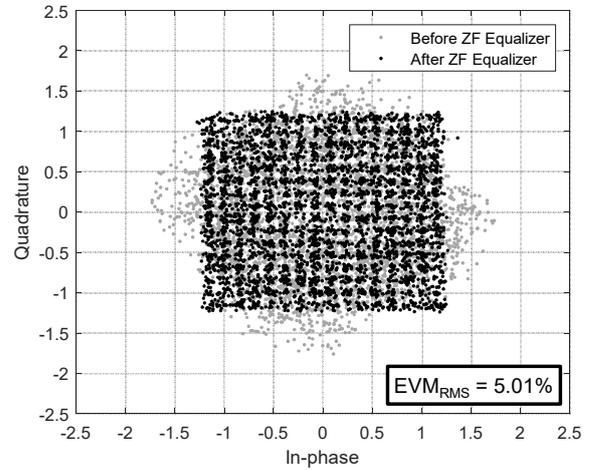

Fig. 8. TRxPs transmitting simultaneously. CJT received symbols constellation and EVM measurement before and after ZF equalization.

## V. CONCLUSIONS

Analog fronthaul is a prospective solution for the development of 6G new services and scenarios. The use of ARoF can address many of the challenges posed by 6G, such as the need for lower power consumption, precise synchronization, and large bandwidths. Additionally, ARoF can enable compact radio devices and centralized signal processing allowing for greater flexibility in network design and deployment which can facilitate D-MIMO networks.

In this paper, we assessed the feasibility of CJT in D-MIMO networks enabled by ARoF. We experimentally validated that centralized processing in combination with analog fronthaul links can fulfill the synchronization requirements of CJT. In our tests, channel estimation, time alignment, and precoding are executed in a two TRxPs D-MIMO network achieving a CJT gain of ~5.3 dB (close to the theoretical gain of 6 dB). It is noted that these results are in line with 3GPP conformance tests methods, e.g., OFDM waveform transmission and ZF equalization [11].

Furthermore, to the best of our knowledge, this is the first time that CJT is realized where an ARoF fronthaul link is used and the delay difference between fronthaul links is larger than the signal CP length which is expected to happen in D-MIMO systems.

## ACKNOWLEDGMENT

This project is financially supported by the Swedish Foundation for Strategic Research (project No. SM21-0047). Also, it is supported in part by the Swedish Research Council (VR) project 2019-05197 and Vinnova project 2022-02545 (A-FRONTHAUL).